\documentclass[letterpaper, 10 pt, conference]{ieeeconf}  

\IEEEoverridecommandlockouts                              

\usepackage{epstopdf}
\usepackage{mathptmx}
\usepackage{times}
\usepackage{amsmath}
\usepackage{amssymb}
\usepackage{graphicx}
\usepackage{subcaption}
\usepackage[ruled,vlined]{algorithm2e}
\usepackage{mathtools}
\usepackage{hyperref}
\usepackage[dvipsnames]{xcolor}
\usepackage{caption}
\usepackage{amsmath}

\usepackage{amsthm}

\newtheoremstyle{plainupright}
  {1pt} 
  {1pt} 
  {\normalfont} 
  {} 
  {\bfseries} 
  {.} 
  {.5em} 
  {} 

\theoremstyle{definition}
\newtheorem{definition}{Definition}
\newtheorem{assumption}{Assumption}
\theoremstyle{plain}
\newtheorem{theorem}{Theorem}

\theoremstyle{plainupright}
\newtheorem{remark}{Remark} 

\title{\LARGE \bf
	Adaptive Disturbance Observer-based Full-Order Integral-Terminal Sliding Mode Control with Unknown A Priori Bound on Uncertainty}

\author{Jit Koley$^{1}$,
	and Binoy Krishna Roy$^{2}$, \emph{Senior Member, IEEE}
	\thanks{$^{1}$ Department of Electrical Engineering, IIT Bombay, Mumbai.
		{Email: jit.koley94@gmail.com, $^{2}$Department of Electrical Engineering, National Institute of Technology Silchar, Silchar.
		{Email: bkr@ee.nits.ac.in}.}}%
}
\begin{document}

\maketitle
\thispagestyle{empty}
\pagestyle{empty}

\begin{abstract}
This study presents a novel, continuous finite-time control strategy for a class of nonlinear systems subject to matched uncertainties with unknown bounds. We propose an Adaptive Disturbance Observer-based Full-order Integral-Terminal Sliding Mode Control (ADO-FOITSMC) to stabilize a chain of integrators in presence of exogenous disturbances whose time derivative is bounded by a constant that is not known a priori. Key features of this approach include a significant reduction in control input chattering and a non-monotonic adaptive law for the observer gains, which prevents overestimation while ensuring the global boundedness of system states. The effectiveness and practical viability of the proposed algorithm are demonstrated through its application to the attitude stabilization of a rigid spacecraft.
\end{abstract}


\section{Introduction}
Initiation of control in the midst of uncertainties has been one of the main topics of concern for the past few decades. Sliding Mode Control (SMC) (see  \cite{utkin1999sliding,edwards1998sliding,shtessel2014sliding} and the references therein) is one of the control strategies that drives state trajectories to a predefined manifold in finite time (\emph{reaching phase}) and constrained to evolve in that manifold (\emph{sliding phase}) which makes the time evolution of state trajectories insensitive to external disturbances. Trajectories converge to the origin asymptotically (Conventional sliding mode \cite{shtessel2014sliding, utkin1999sliding}) or in a finite time (Terminal sliding mode \cite{feng2013nonsingular,yang2013continuous,kamal2016continuous,han2023}). However, SMC often involves high-frequency switching, causing undesirable input "chattering" \cite{utkin1999sliding,edwards1998sliding}.
Unlike conventional linear sliding mode, TSMC ensures finite-time convergence, while Integral Sliding Mode (ISM) \cite{pan2017integral,chalanga2013continuous,cui2017extended,xiong2019integral,yu2019} eliminates the reaching phase, ensuring robustness from the outset. These methods typically require prior knowledge of the acting exogenous disturbance, preferably its upper bound, which may be unavailable or time-varying in practice. Our proposed algorithm incorporates this anonymity in the design of the control law, eliminating the reaching phase and significantly reducing chattering in the control input.

Several adaptive sliding mode control (ASMC) algorithms have been developed to address the limitations of traditional SMC. One approach involves monotonically increasing the gain in the signum structure \cite{su1994adaptive,chang2009adaptive,nasiri2014adaptive,negrete2016second}, but this can lead to excessively high gains \cite{plestan2010new}. A novel ASMC technique \cite{roy2020adaptive} eliminates the need for a known disturbance bound, ensuring uniformly ultimate boundedness of state trajectories. Another method \cite{ding2021adaptive} introduces an adaptive second-order SMC using the Lyapunov approach, assuming bounded uncertainties rather than bounded time derivatives. Our proposed methodology further relaxes these constraints, requiring only that the time derivative of disturbances be bounded by an unknown constant, providing a more flexible and robust solution.

Continuous adaptive super-twisting control (STC) algorithms \cite{moreno2012strict,levant2003higher} fall into two categories. The first \cite{plestan2010new,taleb2013pneumatic,alwi2013adaptive,bartolini2013adaptation} increases STC gains until second-order sliding mode (2SM) is achieved, then fixes them. The second \cite{utkin2013adaptive,edwards2014dual,edwards2016adaptive} minimises gains to attain 2SM, reducing chattering and avoiding overestimation. Research on gain minimisation remains limited. Our Lyapunov-based approach ensures global uniform ultimate boundedness (GUUB) of state trajectories and estimation errors, handling disturbances with unknown constant-bounded time derivatives while preventing gain overestimation. We validate its effectiveness through comparisons with existing methods and application to rigid spacecraft attitude stabilisation.
In this manuscript, our main contributions are as follows.
\begin{itemize}
	\item An Adaptive Disturbance Observer-based Full-order Integral-Terminal Sliding Mode Control (ADO-FOITSMC) that eliminates the reaching phase, ensuring robustness from initiation of control.
	\item The proposed algorithm guarantees that the state trajectories are Globally Uniformly Ultimately Bounded (GUUB). This stability is achieved under the relaxed condition that the disturbance's time derivative is bounded, even when the specific value of this bound is unknown a priori.
	\item The proposed adaptive control law avoids gain overestimation while significantly reducing chattering.
\end{itemize}

Throughout this paper, an open ball of radius $r$ at time $t$ is denoted $\mathfrak{B}_r(\mathbf{x}, t) = \{\mathbf{x} \in \mathcal{M} : \|\mathbf{x}\| < r \text{ at time } t\}$. For $\alpha \in \mathbb{R}$ and $x \in \mathbb{R}^n$, $[x]^{\alpha}\text{sgn}(x) = [|x_1|^{\alpha}\text{sign}(x_1), \dots, |x_n|^{\alpha}\text{sign}(x_n)]^T$, with $\text{sgn}(y)$ as the standard signum function for $y \in \mathbb{R}$. The function arguments may be omitted for brevity where the context is clear. 

\section{Problem Formulation}
\label{2}
Consider an $\mathrm{n}^\mathrm{th}$ order chain of integrators which is affine in control and perturbed by matched disturbance
\begin{align}
	\left\{\begin{array}{l}
		\dot{x}_1 =x_2\\
		\dot{x}_2 =x_3\\
		\vdots \\
		\dot{x}_n =f(\textbf{x},t)+b(\textbf{x},t)u(t)+d_{0}\left(\textbf{x},t\right) \label{plant}
	\end{array}\right.
\end{align}
where $\textbf{x}(t)=\left[x_1(t),x_2(t),\cdots,x_n(t)\right]^T \in \mathbb{R}^\mathrm{n}$ comprises all the states in a vector representation, $f:\mathbb{R}^\mathrm{n}\times \mathbb{R}^{\geqslant 0} \to \mathbb{R}$ is the nonlinear drift which is considered to be \emph{locally Lipschitz}, $b:\mathbb{R}^\mathrm{n}\times \mathbb{R}^{\geqslant 0} \to \mathbb{R}$ is the gain for control input, $u(t) \in \mathbb{R}$ is the control input and $d_{0}:\mathbb{R}^\mathrm{n}\times \mathbb{R}^{\geqslant 0} \to \mathbb{R}$ is the unknown perturbation 
representing \textit{nonparametric uncertainties} and \textit{external disturbances}. Further $f(\textbf{x},t)=f_n\left(\textbf{x},t\right)+ f_{\Delta}\left(\textbf{x},t \right)$, 
where $f_n$ is the nominal component of $f$ and $ f_{\Delta}$ is the perturbation from the known nominal value.
\begin{assumption}\label{ass_1}
	The time derivative of parametric perturbation $ f_{\Delta} \left(\textbf{x},t \right)$ and matched uncertainty $d_{0} \left(\textbf{x},t \right)$, are assumed to be globally bounded, i.e. there exists some real positive constants $f_{\mathrm{max}}$ and $d_{\mathrm{max}}$ such that $|\dot{f}_{\Delta}\left(\textbf{x},t \right)| \leqslant f_{\mathrm{max}}$ and $|\dot{d}_{0}\left(\textbf{x},t \right)| \leqslant d_{\mathrm{max}}$ where $f_{\mathrm{max}}, d_{\mathrm{max}} \in \mathbb{R}^{\geqslant 0}$. For the subsequent analysis, these bounds are consolidated into a single constant $k=f_{\rm max}+d_{\rm max}$.
\end{assumption}
\begin{assumption}\label{assum2}
    The specific values of the constant bounds $f_{\mathrm{max}}$ and $d_{\mathrm{max}}$ introduced in Assumption \ref{ass_1} are not known beforehand.
\end{assumption}
In sliding mode control, the system's state trajectory is driven towards a predefined sliding manifold, $s(t)=0$. The behavior of the system once it reaches this manifold (or its vicinity) at a finite time $t_{r}$, known as the reaching time, can be categorized as follows.

\begin{definition}[Ideal sliding \cite{levant2003higher}]
An \emph{ideal sliding} mode represents a scenario where the system's state trajectory perfectly tracks the sliding manifold after reaching it. Rigorously, an ideal sliding mode is said to be established at time $t_{r}$ if for some finite $t_{r} \geqslant t_{0}$, $s(t) \equiv 0, \;\forall t \geq t_{r}$.
\end{definition}
\begin{definition}[Real sliding \cite{levant2003higher}]
	For a given, arbitrarily small constant $\varepsilon >0$, a \emph{real sliding} mode is achieved on the sliding manifold $s(t)=0$ at time $t_{r}\geqslant t_{0}$, if $\vert s(t) \vert \leqslant \varepsilon ,\; \forall t \geqslant t_r$.
\end{definition}
In real-time applications, control algorithms operate with discrete measurements and thereby, introduce imperfections in switching. Thus, it's obdurate to obtain ideal sliding in such cases. However, real sliding can be obtained with a notable relaxation in sliding accuracy and effectively reducing chattering.\par
Consider a full-order integral terminal sliding manifold (FOITSM) of the following form
\begin{equation} \label{ss}
\begin{cases}
    s = s_{0}(\mathbf{x}) - z, \\
    \begin{aligned}
      \dot{z} = & -C_n|x_n|^{\alpha_n}\operatorname{sgn}(x_n) - C_{n-1}|x_{n-1}|^{\alpha_{n-1}}\operatorname{sgn}(x_{n-1}) \\
                & - \cdots - C_1|x_1|^{\alpha_1}\operatorname{sgn}(x_1),
    \end{aligned}
\end{cases}
\end{equation}
where $s_{0}(\textbf{x})=x_n(t)$, $z(0)=s_0(0)$, and $C_i$'s and $\alpha_i$'s are constants which are chosen such that the polynomial $p^n+C_np^{n-1}+C_{n-1}p^{n-2}+\cdots+C_2p+C_1$ is \textit{Hurwitz} stable and $\alpha_i$'s are determined based on the following conditions 
\begin{equation}
	\left\{\begin{array}{l}
		\alpha_1=\alpha, \; \; n=1, \\
		\alpha_{i-1}=\frac{\alpha_i \alpha_{i+1}}{2\alpha_{i+1}-\alpha_i}, \; \; i=1,\cdots, n, \; \; \forall n \geqslant 2,\nonumber
	\end{array}\right.
\end{equation}
where $\alpha_{n+1}=1$, $\alpha_n=\alpha$, $\alpha \in (1- \varepsilon, 1)$, $\varepsilon \in (0,1)$.
\begin{remark}
	The condition \(z(0) = s_0(0)\) ensures \(s = 0\) at \(t = 0\), enabling sliding motion from the outset, eliminating the reaching phase typical in conventional TSMC.
\end{remark}
Our goal is to synthesize a feedback controller for the system (\ref{plant}) subject to the disturbances as characterised in Assumptions \ref{ass_1} and \ref{assum2}. The goal is to steer the system's trajectories to, and maintain them within, a bounded vicinity of the sliding manifold (\ref{ss}), thereby realizing a \emph{real sliding} mode.

\section{Chattering free adaptive SMC with unknown apriori bound on uncertainty} \label{sec4}
This section details the design of a feedback control law for system (\ref{plant}) following the assumptions on parametric uncertainties and external disturbances outlined in \ref{ass_1} and \ref{assum2}. 
\begin{theorem}\label{foitadoTH}
	The system \eqref{plant}, characterised by disturbance $d$ as assumed in \ref{ass_1} and \ref{assum2}, is globally uniformly stable (GUB) and globally uniformly ultimately bounded (GUUB) (\cite{khalil2002nonlinear}) in  finite time $t_{r}$ 
	\begin{align}
		t_{r}& \leqslant \frac{1}{\gamma}\text{ln}\left[ \frac{V(0)-\frac{\bar{\delta}}{\gamma}}{\bar{\delta}\left( \frac{1}{\gamma-\theta}-\frac{1}{\gamma}\right)}\right],
	\end{align}
	if the control $u$ is chosen as
	\begin{align}
		u=&b^{-1}(\textbf{x},t)(u_{\text{eqv}}+u_{\text{ado}}),\label{uado}\\
		u_{\text{eqv}}=& -f_n(\textbf{x},t)-C_n|{x_n}|^{\alpha_n}\text{sgn}(x_n) \nonumber \\
		&-C_{n-1}|{x_{n-1}}|^{\alpha_{n-1}}\text{sgn}(x_{n-1})-...-C_1|{x_1}|^{\alpha_1}\text{sgn}(x_1), \nonumber\\
		u_{\text{ado}}=& -\kappa s-\hat{d}.\label{conado}
	\end{align} 
        The disturbance observer dynamics are defined as
\begin{equation} \label{eq:observer_dynamics}
\begin{cases}
    \hat{d} = \lambda (x_n - \zeta), \\
    \dot{\zeta} = f_n(\mathbf{x}, t) + b(\mathbf{x}, t)u + \hat{d} - \frac{\hat{k}}{\lambda} \operatorname{sgn}(\tilde{d}) - \frac{s}{\lambda},
\end{cases}
\end{equation}
with the adaptation law
\begin{align}
	\dot{\hat{k}} &= -\tau \hat{k} + \mu |s|, \quad \hat{k}(0) > 0, \label{adalado}
\end{align}
where  \( \tilde{d} = d - \hat{d} \), \( \tilde{k} = k - \hat{k} \), \( z(0) = x_n(0) \), \( s(0) = s_0 \), \( \tau = \mu + 1 + \tau_0 \), and \( \lambda, \mu, \tau_0, \kappa \in \mathbb{R}^+ \). For \( \mathcal{S} = [s, \tilde{d}, \tilde{k}]^T \), the ultimate bound on \( \|\mathcal{S}\|^2 \) is:
\begin{align}
	\mathcal{B} = \sqrt{\frac{2 \bar{\delta}}{\gamma - \theta}},
\end{align}
where \( \bar{\delta} = \frac{1}{2} \tau k^2 \) and $\lVert \cdot \rVert$ denotes the usual Euclidean norm.
\end{theorem}

\begin{proof}
    Consider a Lyapunov function candidate as
\begin{align}
	V =& \frac{1}{2}s^2+\frac{1}{2}\tilde{d}^2+\frac{1}{2}\tilde{k}^2, \nonumber \\
	\dot{V}=& -\kappa s^{2}+s\tilde{d}+\tilde{d}\left (\dot{d}-\lambda \tilde{d}-\hat{k}\text{sgn}(\tilde{d})-s\right)+\tilde{k}(-\dot{\hat{k}}),\nonumber\\
	\leqslant & -\kappa s^{2}-\lambda \tilde{d}^{2}+\frac{1}{2}\tilde{k}^{2}+\frac{1}{2}\tilde{d}^{2}+\tilde{k}\left(\tau \hat{k}-\mu |s|\right),\nonumber\\
	=& -\kappa s^{2}-\left (\lambda -\frac{1}{2}\right )\tilde{d}^{2}+\frac{1}{2}\tilde{k}^2+\tilde{k}\left(\tau \hat{k}-\mu |s|\right),\nonumber\\
	\leqslant& -\kappa s^{2}-\left (\lambda -\frac{1}{2}\right )\tilde{d}^{2}+\frac{1}{2}\tilde{k}^2+\tau \tilde{k}\hat{k}+\frac{1}{2} \mu \tilde{k}^{2}+\frac{1}{2} \mu s^{2},\nonumber\\
	=& -\left (\kappa-\frac{1}{2} \mu \right)s^2-\left (\lambda -\frac{1}{2}\right )\tilde{d}^{2}+\frac{1}{2}(\mu+1)\tilde{k}^{2}\nonumber\\
	&+\tau \left(k \hat{k}-\hat{k}^{2} \right). \label{pf_1}
\end{align}
Using the fact that $k\hat{k} \geqslant 0$ for all time and,
\begin{align}
	k\hat{k}-\hat{k}^2 &=-\left( \frac{k}{\sqrt{2}}-\frac{\hat{k}}{\sqrt{2}} \right)^{2}-\frac{\hat{k}^2}{2}+\frac{k^2}{2}\leqslant -\frac{1}{2}\tilde{k}^{2}+\frac{k^2}{2}\label{ineq_2}.
\end{align}
Using (\ref{ineq_2}), (\ref{pf_1}) becomes
\begin{align}
	\dot{V} \leqslant & -\left (\kappa-\frac{1}{2} \mu \right)s^2-\left (\lambda -\frac{1}{2}\right )\tilde{d}^{2}-\frac{1}{2} \left (\tau- \mu -1 \right)\tilde{k}^2+\frac{1}{2}\tau k^2, \nonumber\\
	    \leqslant & -\gamma V+\Bar{\delta},\nonumber
\end{align}
where $\gamma = \text{min} \left \{\Bar{\kappa},\Bar{\lambda}, \frac{\tau_{0}}{2} \right\}>0$, $\Bar{\kappa}=\kappa-\frac{1}{2}\mu$, $\Bar{\lambda}=\lambda-\frac{1}{2}$, $\tau_{0}=\left(\tau-\mu-1 \right)$ and $\bar{\delta}=\frac{1}{2} \tau k^{2}$. \par
For $0<\theta<\gamma$, the above equation can be rewritten as,
\begin{align}
	\dot{V}&=-\theta V-\left( \gamma - \theta \right)V+\Bar{\delta} \nonumber.
\end{align}
For some $\mathcal{R}=\frac{\Bar{\delta}}{\gamma - \theta}$, $\dot{V}(t) < 0$ for $V(0)> \mathcal{R}$. Hence $V(t)$ reaches $\mathcal{R}$ in finite time, say $t_{r}$ and remains there in the closed ball of radius $\mathcal{R}, \; \forall t >t_{r}$, where $t_{r}$ can be evaluated as given below
\begin{align}
	t_{r}& \leqslant \frac{1}{\gamma}\text{ln}\left[ \frac{V(0)-\frac{\bar{\delta}}{\gamma}}{\bar{\delta}\left( \frac{1}{\gamma-\theta}-\frac{1}{\gamma}\right)}\right].
\end{align}
As the term containing sliding variable $\frac{1}{2}s^{2} \leqslant V $, the state trajectories are also confined within the closed ball of radius $\sqrt{\frac{2 \Bar{\delta}}{\gamma - \theta}}$. This ends the proof. \qed
\end{proof}

\begin{remark}
	The governing disturbance observer dynamics in \eqref{eq:observer_dynamics}, a dynamical equation in $\zeta$, is continuous since the discontinuity lies in $\dot{\zeta}$.
\end{remark}

\begin{figure}[!h]
	\centering
	\includegraphics[width=0.5\textwidth,keepaspectratio]{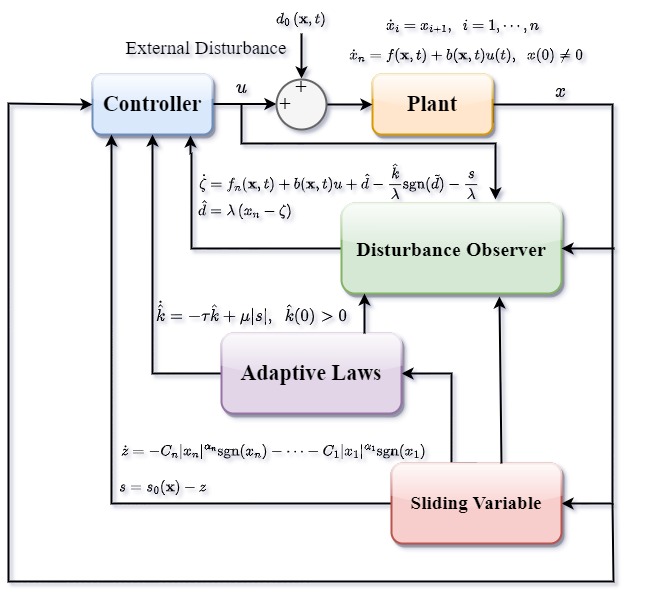}
	\caption{\small Schematic Diagram for Stabilisation of (\ref{plant})}
	\label{sche_pl}
\end{figure}

\begin{remark}
	A slight modification to the control law (\ref{conado}) as
	\begin{align}
		u_{\text{f}_\text{ado}}=-\kappa_{1}s -\kappa_{2}|s|^{\frac{1}{2}}\text{sgn}(s), \nonumber 
	\end{align}
	for $\kappa_{1}>\frac{1}{2}\mu$, $\kappa_{2} >0$ would result in a Fast FOITSMC with a faster convergence rate.
\end{remark}

\begin{remark}
	The Lyapunov stability analysis of the closed-loop system distinctively incorporates disturbance error dynamics, which sets it apart from other existing disturbance observer designs in the literature. The schematic in fig. \ref{sche_pl} illustrates the control architecture.
\end{remark}

\begin{remark}
	In \eqref{eq:observer_dynamics}, the term \( \text{sgn}(\tilde{d}) \) is used, but since the actual disturbance \( d \) is unmeasurable, we define:
	\begin{align}
    \omega(t) = \int_{t_0}^t \tilde{d}(\sigma) \, d\sigma 
    =& x_n - \int_{t_0}^t \left( f(\mathbf{x}, \sigma) + b(\mathbf{x}, \sigma)u(\sigma) \right. \nonumber\\
    &\left. \quad + \hat{d}(\mathbf{x}, \sigma) \right) \, d\sigma,
    \end{align}
	and define \( \text{sgn}(\tilde{d}(t)) = \text{sgn}(\omega(t) - \omega(t - \tau_d)) \), where \( \tau_d \) is the system's sampling time. Since \( \tilde{d} = \lim_{\tau_d \to 0} \frac{\omega(t) - \omega(t - \tau_d)}{\tau_d} \), only the sign of \( \tilde{d} \) is needed, which is simpler to compute.
\end{remark}

\begin{remark}
	The dynamical equation \eqref{eq:observer_dynamics} constitute the disturbance observer for system \eqref{plant} with the characterisation of disturbance $d$ as given in assumptions \ref{ass_1} and \ref{assum2}.
\end{remark}
\section{Illustrative Example I} \label{sec5}
Consider a $3^{\text{rd}}$ order chain of integrator as
\begin{equation}
	\left\{\begin{array}{l}
		\dot{x}_1=x_2\\
		\dot{x}_2=x_3\\
		\dot{x}_3=u+d
		\label{pado3}
	\end{array}\right.
\end{equation}
where the disturbance $d=\text{sin}(2\pi t)$. The sliding surface is chosen as follows
\begin{align}
	s&=x_3-z, \nonumber \\
	\dot{z}&=15\;\text{sgn}(x_3)\vert x_3 \vert^{\frac{7}{10}}+66\;\text{sgn}(x_2) \vert x_2 \vert^{\frac{7}{13}}+80\;\text{sgn}(x_1) \vert x_1 \vert^{\frac{7}{16}}. \label{stcss}
\end{align}
The control input $u$ is designed as,
\begin{align}
	u=&-15\text{sgn}(x_3)\vert x_3 \vert^{\frac{7}{10}}-66\text{sgn}(x_2) \vert x_2 \vert^{\frac{7}{13}} \nonumber\\
	&-80\text{sgn}(x_1) \vert x_1 \vert^{\frac{7}{16}}+u_{ado} \label{u_ado3}
\end{align}
and $u_{ado}$ is designed as in (\ref{conado}), (\ref{eq:observer_dynamics}) and (\ref{adalado}) with parameters $\kappa=5$, $\lambda=5$, $\tau=5$ and $\mu=2$.
\begin{figure}[h!]
    \centering
    \subfloat[States with adaptive ADO based FOITSMC\label{ado3_states}]{%
        \includegraphics[width=0.23\textwidth]{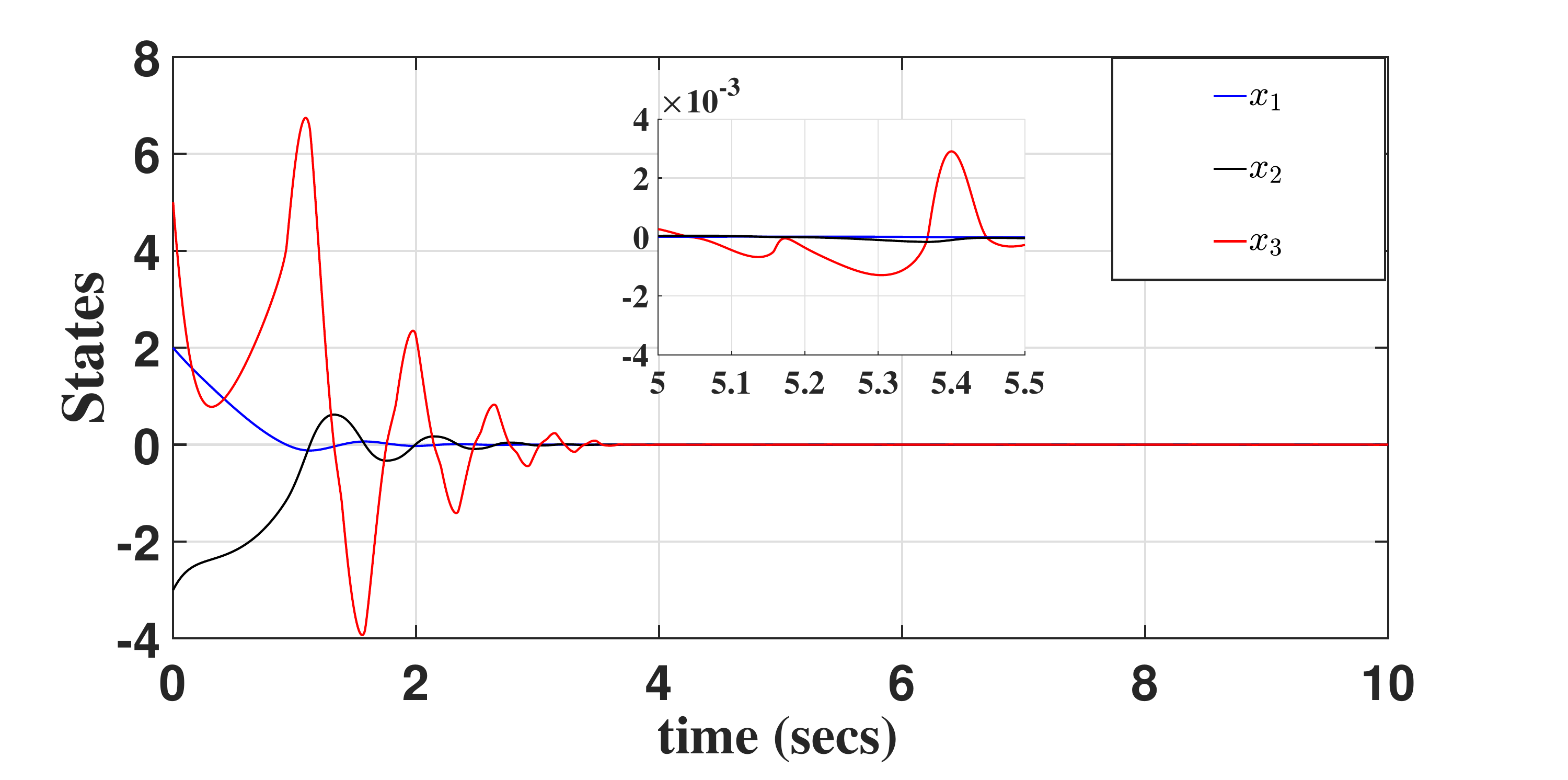}}
    \hfill 
    \subfloat[Sliding variable of adaptive ADO based FOITSMC\label{ado3_ss}]{%
        \includegraphics[width=0.23\textwidth]{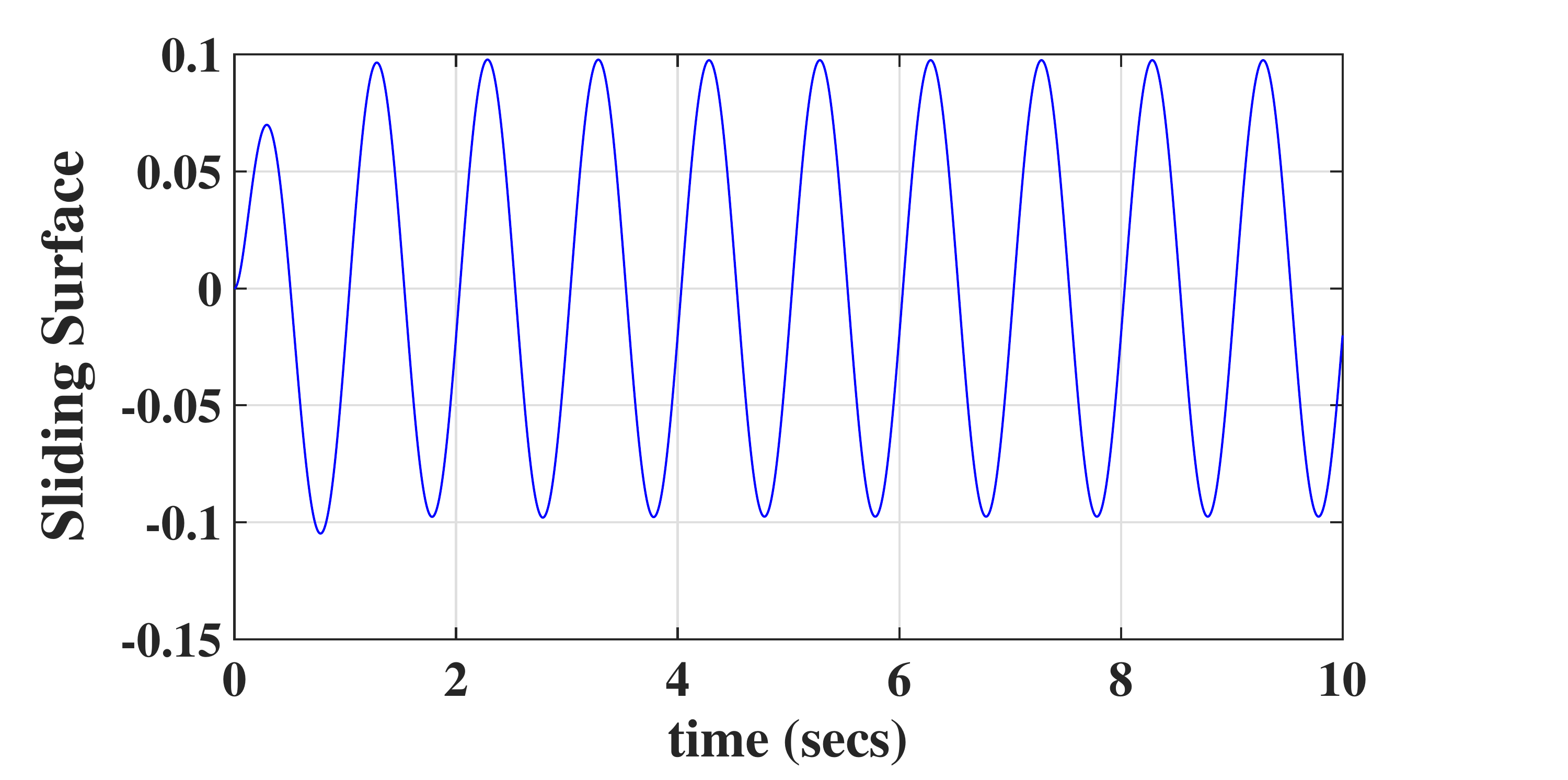}}
    \\ 
    \subfloat[Input for FOITSMC with adaptive ADO\label{ado3_u}]{%
        \includegraphics[width=0.23\textwidth]{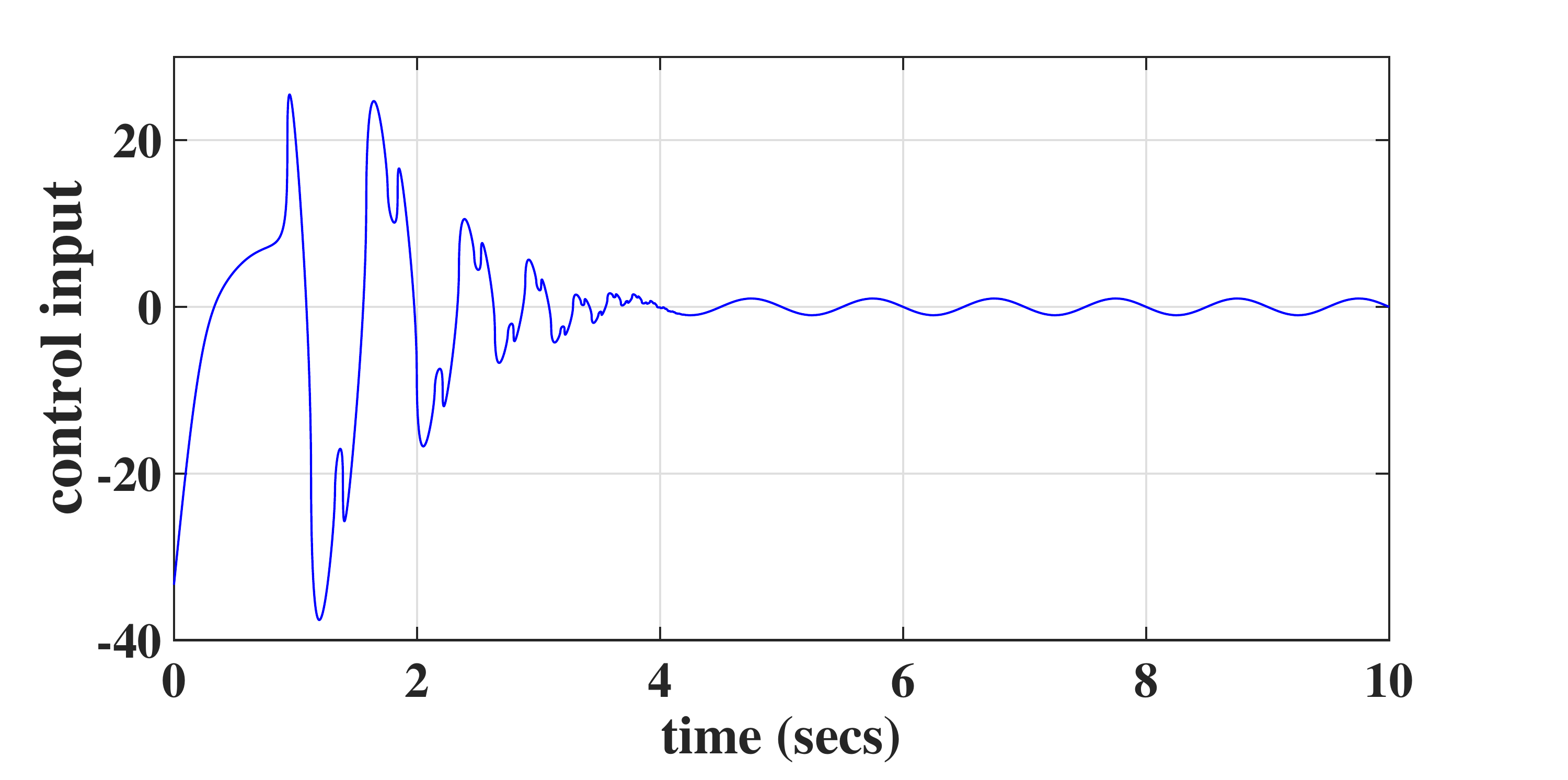}}
    \hfill 
    \subfloat[Disturbance estimation error for adaptive ADO based FOITSMC\label{ado3_d_tilde}]{%
        \includegraphics[width=0.23\textwidth]{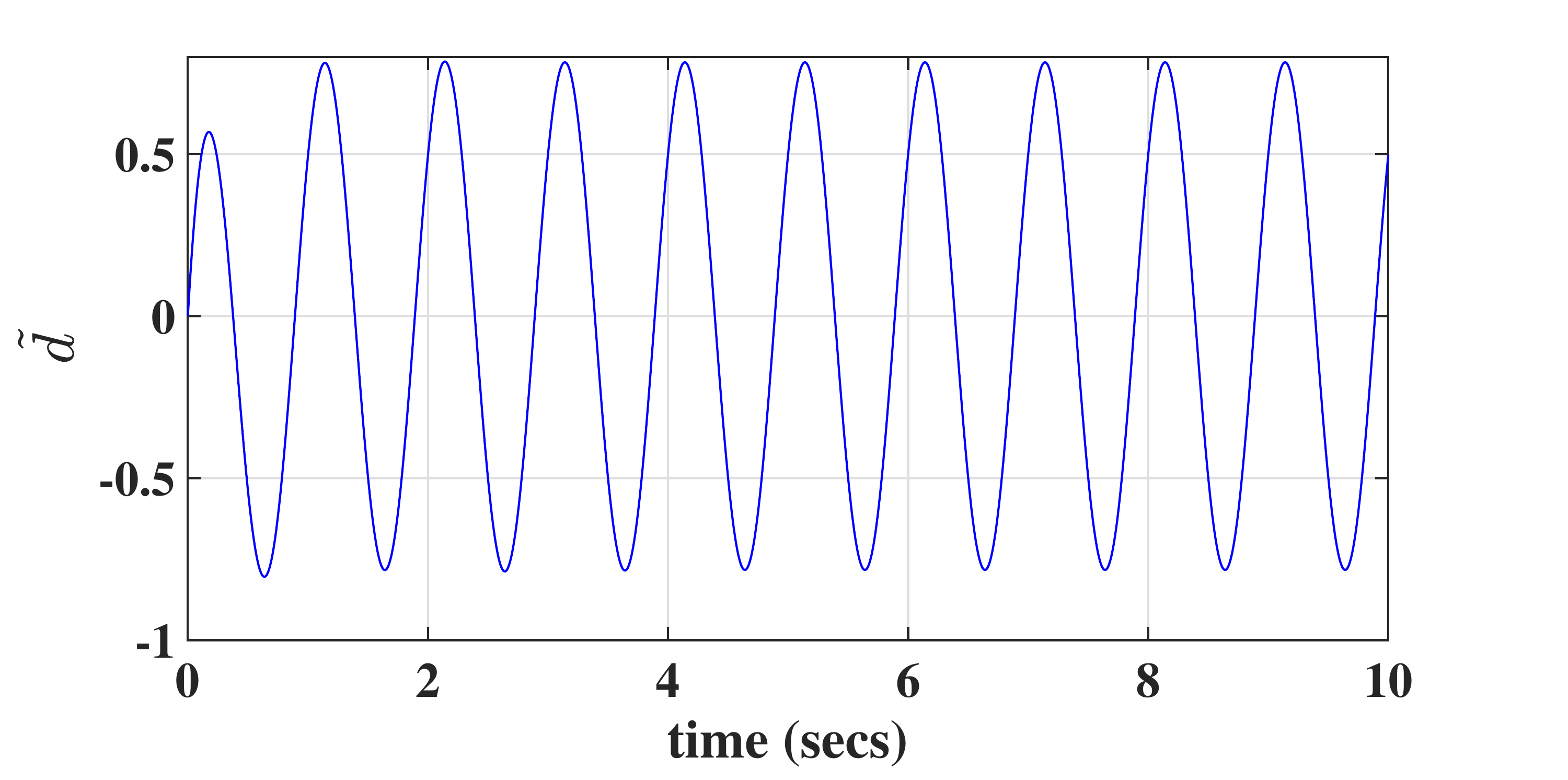}}
    \\
    \subfloat[Estimation of $k$ for FOITSMC with adaptive ADO\label{ado3_k_hat}]{%
        \includegraphics[width=4cm, height=3cm, keepaspectratio]{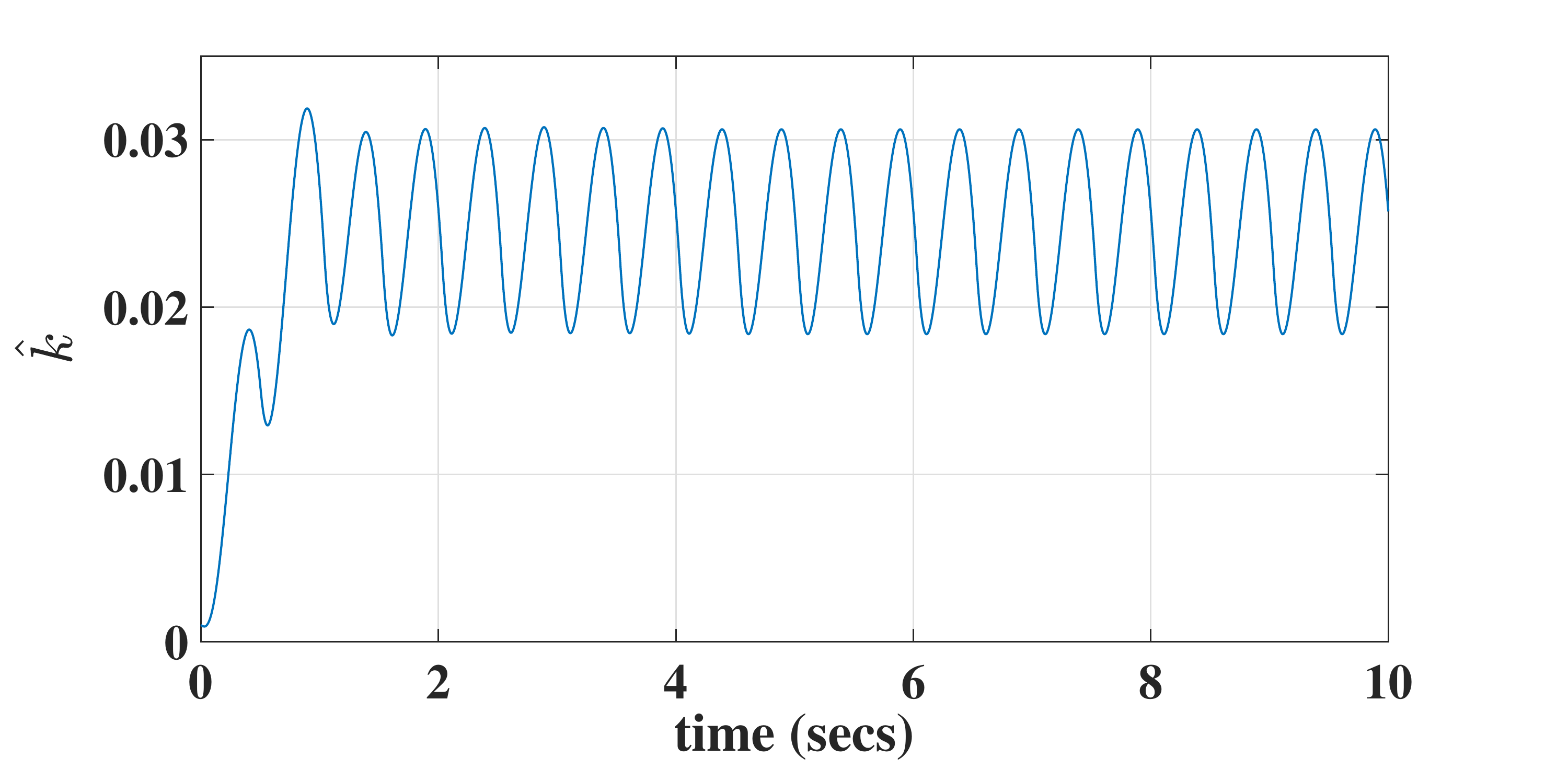}}
        
    \caption{\small \ref{ado3_states}, \ref{ado3_ss} and \ref{ado3_d_tilde} show the ultimate boundedness of state trajectories, sliding variable and disturbance estimation error, respectively, on the application of adaptive ADO-based FOITSMC. Moreover, the minimum estimated adaptive gain required for GUUB is shown in \ref{ado3_k_hat}.}
    \label{fig_ado} 
\end{figure}

\begin{remark}
	Adaptive super-twisting control (STC) strategies \cite{plestan2010new,taleb2013pneumatic,utkin2013adaptive,edwards2014dual} often use signum functions in adaptation laws, which can lead to over- or under-estimated gains. This occurs when gains increase despite a decreasing sliding variable magnitude, or vice versa. In contrast, the adaptive law in \eqref{adalado} is devoid of any such discontinuous functions, thus striving \(\hat{k}\) to achieve the desired performance with lesser gain values.
\end{remark}

Simulation results were generated using MATLAB R2021a’s ODE4 (Runge-Kutta) solver with a fixed step size of 0.001. Figures \ref{ado3_states} and \ref{ado3_ss} show that the states of \eqref{pado3} and the sliding variable, respectively, remain within specified bounds in finite time under the control input \eqref{u_ado3}. Figure \ref{ado3_u} illustrates a chatter-free control input. The estimated $k$ is non-monotonic (as shown in fig. \ref{ado3_k_hat}) and it decreases (or increases) depending on whether the magnitude of the sliding variable decreases (or increases).

\section{Performance Comparison} \label{sec6}
To demonstrate the effectiveness of the proposed algorithm, we present a performance comparison with existing methods. For this purpose, we consider the Adaptive Super-Twisting control law (ASTW) outlined by (\cite{shtessel2012novel})for reference.
\begin{theorem}[\cite{shtessel2012novel}] Consider a system of the following form
	\begin{align}
		\dot{\sigma}(t)&=w+a_{1}(x,t)
	\end{align}
	such that $\vert \dot{a}_{1}(x,t) \vert<\delta<\infty$, $\delta$ unknown.
	For any given arbitrary initial conditions $x(0)$, $\sigma(0)$, there exists a finite $t_{f} >0$ such that a real $2-$sliding mode has been established $\forall \; t>t_{f}$ via control law 
	\begin{equation}
		\left\{\begin{array}{l}
			\omega = -\alpha \vert \sigma \vert ^{\frac{1}{2}} \text{sgn}(\sigma)+v \\
			\dot{v} =-\frac{\beta}{2}\text{sgn}(\sigma)
		\end{array}\right.
	\end{equation}
	with adaptive gains $\alpha$, $\beta$ defined as follows
	\begin{align}
		\dot{\alpha} &= \begin{cases}
			\omega_{1}\sqrt{\frac{\gamma_{1}}{2}}\text{sgn}\left( \vert \sigma\vert - \mu \right), \; \; &\alpha >\alpha_{m}\\
			\eta, \; &\alpha \leqslant \alpha_{m}
		\end{cases}\\
		\beta &=2\varepsilon \alpha
	\end{align}
	where $\varepsilon$, $\lambda$, $\gamma_{1}$, $\omega_{1}$, $\eta$ are positive constants. The parameter $\alpha_{m}$ is an arbitrary-small constant.
\end{theorem}
\begin{figure}[h!]
    \centering
    \subfloat[The adaptive gain $(\hat{k})$ for disturbance $d=\sin(2\pi t)$\label{com_khat1}]{%
        \includegraphics[width=0.23\textwidth,keepaspectratio]{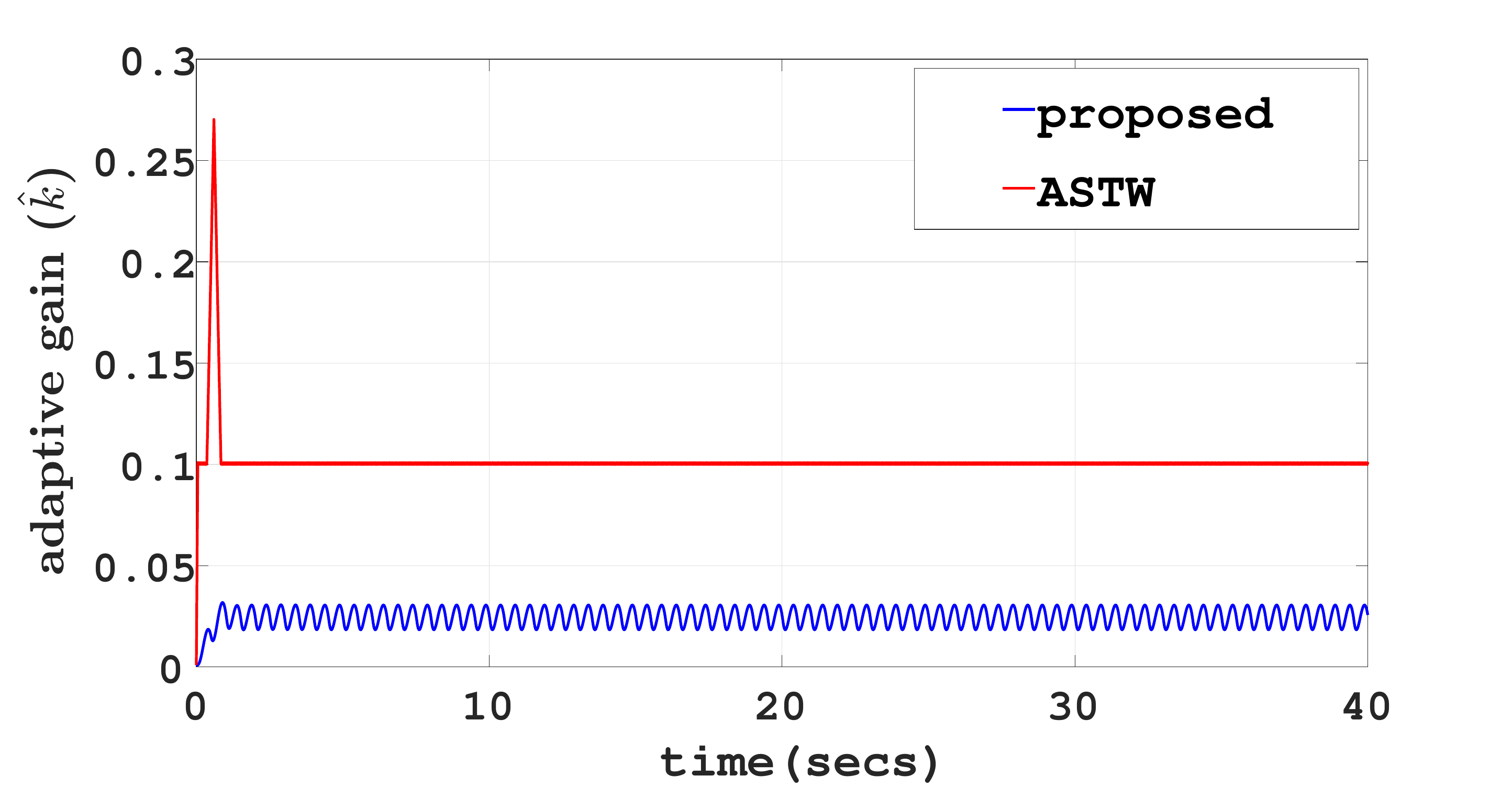}}
    \hfill 
    \subfloat[The adaptive gain $(\hat{k})$ for disturbance $d=\sin(2\pi t)+\texttt{ramp}(t)$\label{com_khat2}]{%
        \includegraphics[width=0.23\textwidth,keepaspectratio]{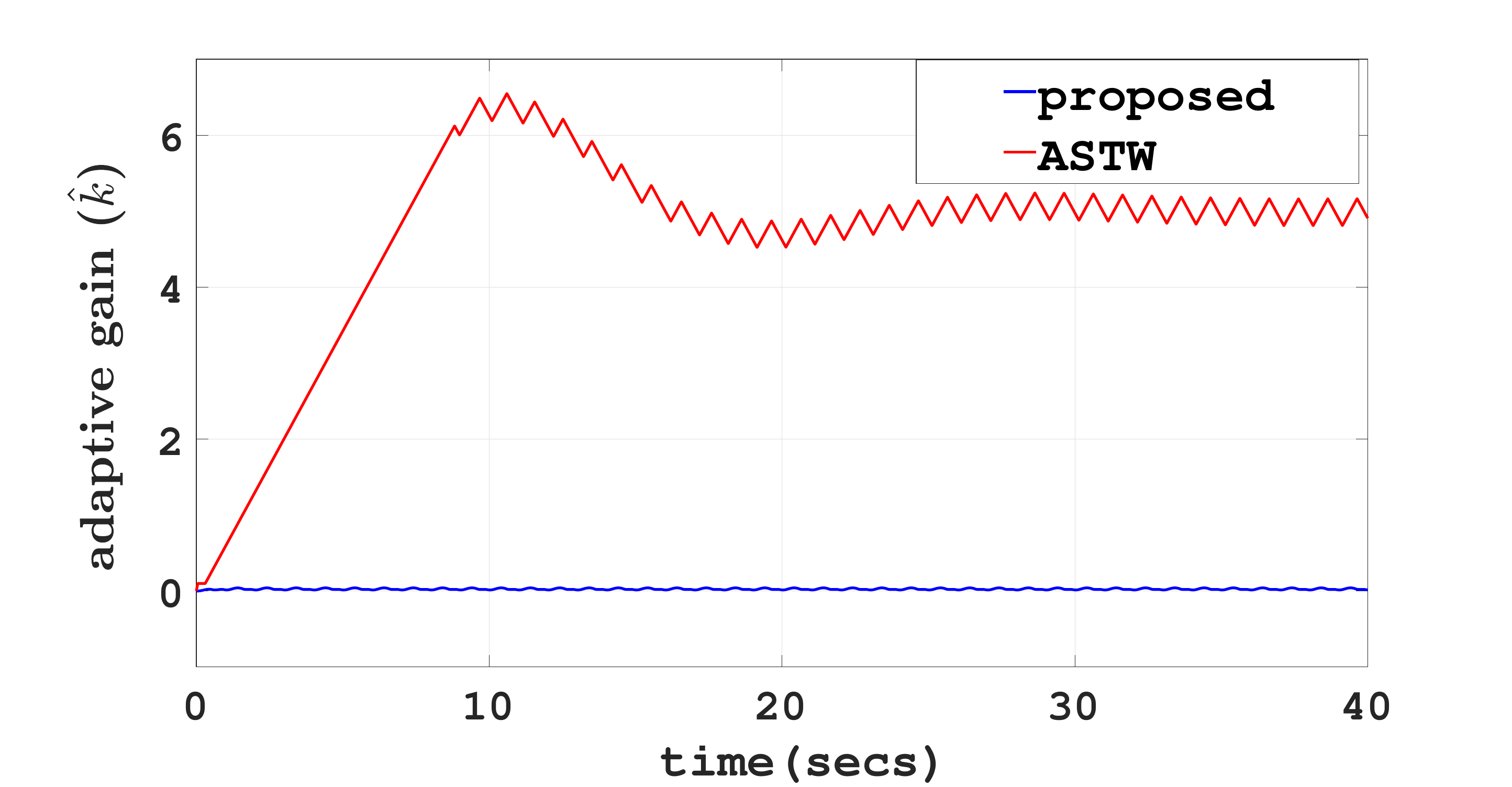}}
    \\ 
    \subfloat[The sliding variable $(s)$ in presence of disturbance $d=\sin(2\pi t)$\label{com_ss1}]{%
        \includegraphics[width=0.23\textwidth,keepaspectratio]{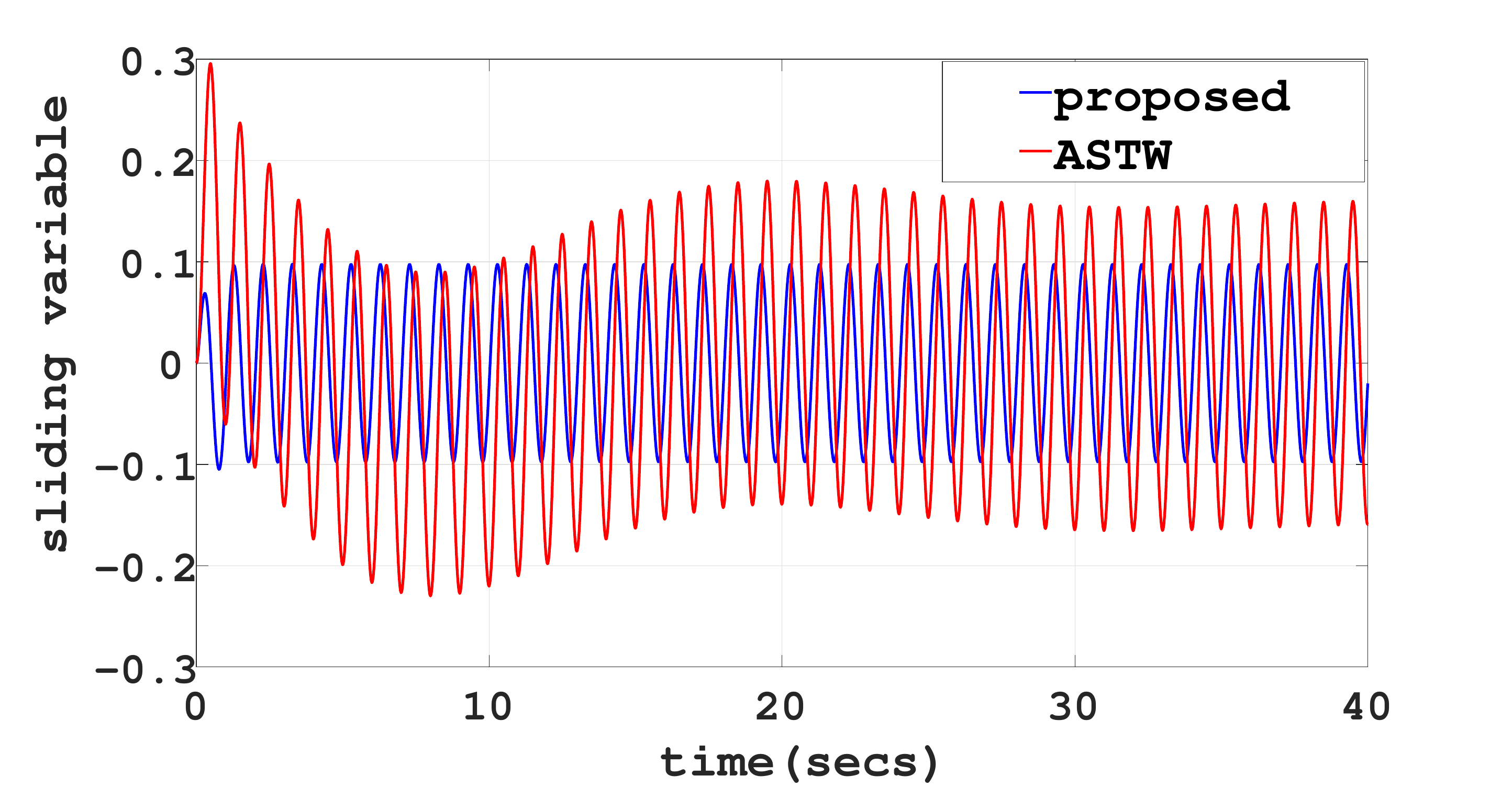}}
    \hfill 
    \subfloat[The sliding variable $(s)$ in presence of disturbance $d=\sin(2\pi t)+\texttt{ramp}(t)$ \label{com_ss2}]{%
        \includegraphics[width=0.23\textwidth,keepaspectratio]{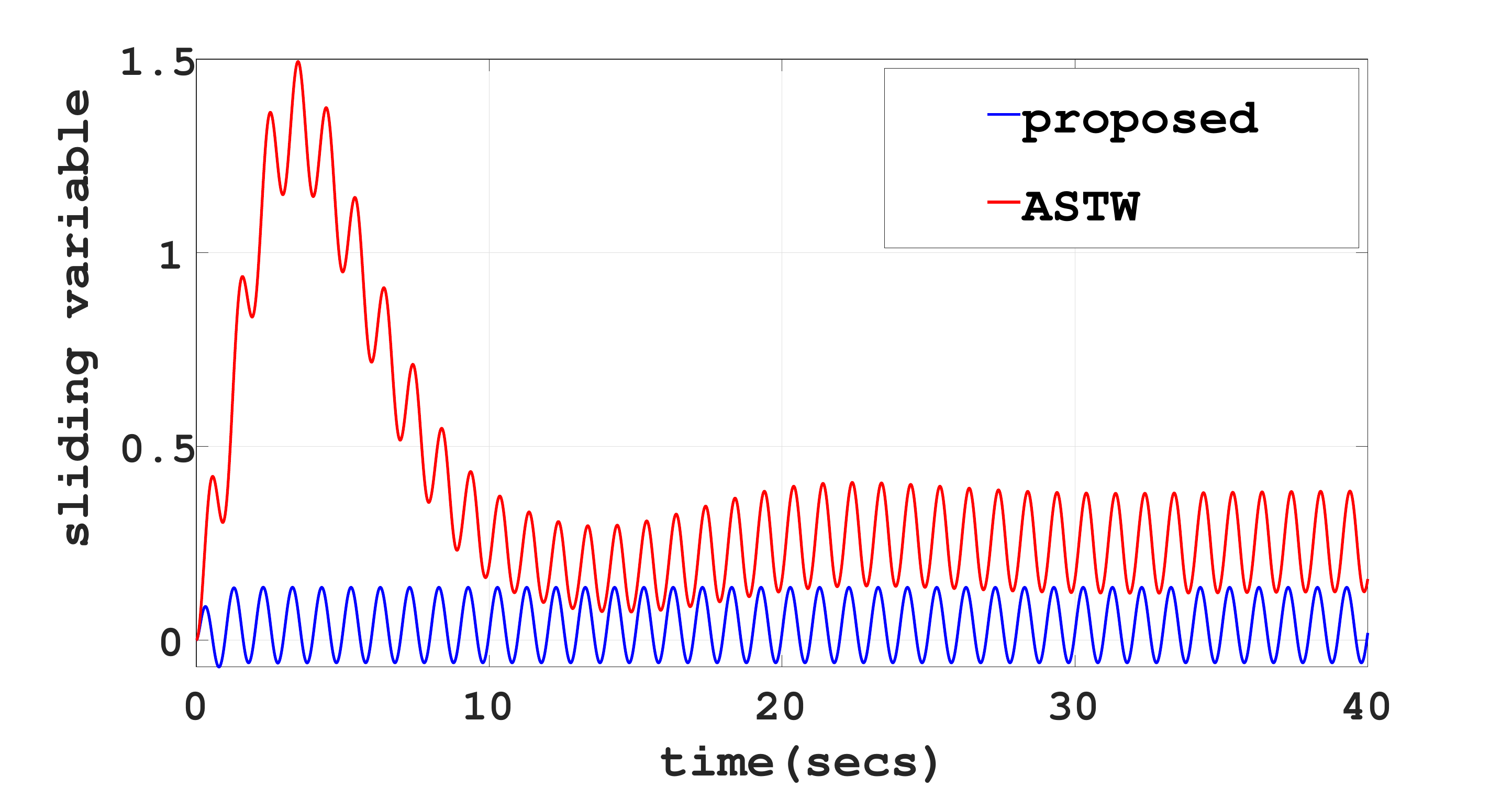}}
        
    \caption{\small The comparison result of the proposed algorithm with the existing ASTW in the presence of disturbances $\sin(2\pi t)$ and $\sin(2\pi t)+\texttt{ramp}(t)$, where $\texttt{ramp}(t)=0.5 t\left(\text{sgn(t)}+1\right)$. \ref{com_khat1} and \ref{com_khat2} shows the comparison of estimated $k$ for both the cases. \ref{com_ss1} and \ref{com_ss2} illustrates the comparison in sliding variable for two kind of disturbances.}
    \label{fig_com} 
\end{figure}
Simulations were conducted using MATLAB R2021a’s ODE4 (Runge-Kutta) solver with a fixed step size of 0.001. The disturbances \( \sin(2\pi t) \) and \( \sin(2\pi t) + \text{ramp}(t) \), where \( \text{ramp}(t) = 0.5 t (\text{sgn}(t) + 1) \), were deliberately chosen to test robustness against bounded and unbounded (with bounded time-derivative) disturbances. Figures \ref{com_khat1} and \ref{com_khat2} show that the proposed algorithm’s estimated gain \(\hat{k}\) remains lower than that of ASTW in both cases. Figures \ref{com_ss1} and \ref{com_ss2} demonstrate that the sliding variable \( s \) in the proposed algorithm stays closer to zero, in comparison to the existing ASTW, signifies greater robustness against the acting disturbances.

\section{Case study} \label{sec7}
Attitude control and stabilisation of a rigid spacecraft under external disturbances have long been a focus of control research \cite{zhu2011adaptive,show2003lmi,luo2005inverse,shen2014integral,xia2019finite,xie2022high} due to the system's highly nonlinear dynamics, which pose significant challenges for control design. This section applies the proposed control algorithm to achieve robust attitude stabilisation of a rigid spacecraft. The unit-quaternion-based attitude dynamics of a rigid spacecraft are described by:
\begin{align}
	\begin{cases}
		\dot{q}_0 &= -\frac{1}{2} q_v^T \Omega, \\
		\dot{q}_v &= \frac{1}{2} (q_0 I_3 + q_v^\times) \Omega, \\
		J \dot{\Omega} &= -\Omega^\times J \Omega + u + d,
	\end{cases}
\end{align}
where the unit quaternion \( q = [q_0, q_1, q_2, q_3]^T = [q_0, q_v]^T \) satisfies \( q_0^2 + q_v^T q_v = 1 \), with \( q_v \in \mathbb{R}^3 \) as the vector part and \( q_0 \in \mathbb{R} \) as the scalar component. Here, \( \Omega \in \mathbb{R}^3 \) is the angular velocity, \( u \in \mathbb{R}^3 \) is the control torque, \( J \in \mathbb{R}^{3 \times 3} \) is the symmetric positive definite inertia matrix, \( I_3 \) is the \( 3 \times 3 \) identity matrix, and the skew-symmetric operator for any vector \( a = [a_1, a_2, a_3]^T \) is:
\begin{align}
	a^\times = \begin{pmatrix}
		0 & -a_3 & a_2 \\
		a_3 & 0 & -a_1 \\
		-a_2 & a_1 & 0
	\end{pmatrix}.
\end{align}
\begin{assumption}
	The inertia matrix $J$ is assumed to be known.
\end{assumption}
\begin{assumption}
	The exogenous matched disturbance $d$ is continuously differentiable such that each components $d_{i}$ of $d$ satisfy, $\dot{d}_{i}\leqslant k_{i}, \; \; \text{for} \; i=1,2,3$ where $k_i$s are finite.
\end{assumption}
\par
Consider a sliding surface of the following form,
\begin{align}
	s=e-e_{0}+\int_{t_{0}}^{t}[e]^{\frac{1}{2}}\text{sgn}(e)\mathrm{d}\tau,
	\nonumber
\end{align}
where $e=\Omega+k_{v}q_{v}$, $k_v >0$. The control input is chosen as follows
\begin{align}
	\nonumber
	u=&\Omega^{\times} J \Omega -\frac{1}{2}k_{v}J\left (q_{0}I_{3}-q_{v}^{\times} \right )\Omega 
	-J[ e]^{\frac{1}{2}}\text{sgn}(e)-J\Theta s\\
	\nonumber
	\hat{d}=&\Lambda \left(\Omega-z \right)\\
	\dot{z}=&-J^{-1}\Omega^{\times}J\Omega +J^{-1}u+J^{-1}\hat{d} \nonumber\\
	&-\Lambda^{-1}[\hat{k}]\text{sgn}(\tilde{d})-\Lambda^{-1}J^{-1}s \nonumber
\end{align}
and adaptation law as
\begin{align}
	\dot{\hat{k}}=&-\tau \hat{k}+\mu |s|, \; \; \hat{k}(0) >0,
	\nonumber
\end{align}
where $\tau$ is a positive definite diagonal matrix with its diagonal elements as $\text{diag}(\tau)=[\tau_{1}, \tau_{2}, \tau_{3}]$ and $\mu$, $\Lambda$ and $\Theta$ are positive definite matrices.

For stability analysis, we consider a Lyapunov function candidate as
\begin{align}
	V=&\frac{1}{2}s^{T}s+\frac{1}{2}\tilde{d}^{T}\tilde{d}+\frac{1}{2}\tilde{k}^{T}\tilde{k} \nonumber\\
	\implies \dot{V}=& s^{T} \left(J^{-1}\tilde{d}-\Theta s \right)+\tilde{d}^{T}\dot{d}-\tilde{d}^{T} \Lambda J^{-1} \tilde{d}-\tilde{d}^{T}\hat{k}\text{sgn}(\tilde{d}) \nonumber\\
	&-\tilde{d}^{T}J^{-1}s-\tilde{k}^{T}\dot{\hat{k}}\nonumber\\
	\leqslant& -s^{T}\Theta s+|\tilde{d}^{i}|k_{i}-\tilde{d}^{T}\Lambda J^{-1} \tilde{d}-|\tilde{d}^{i}|\hat{k}_{i}-\tilde{k}^{T}\dot{\hat{k}}\nonumber\\
	=&-s^{T}\Theta s+|\tilde{d}^{i}|\tilde{k_{i}}-\tilde{d}^{T}\Lambda J^{-1} \tilde{d}-\tilde{k}^{T}\dot{\hat{k}}\nonumber\\
    \leqslant & -s^{T}\Theta s+\frac{1}{2}\tilde{d}^{T}\tilde{d}+\frac{1}{2}\tilde{k}^{T}\tilde{k}-\tilde{d}^{T}\Lambda J^{-1} \tilde{d}\nonumber\\
	&+\tilde{k}^{T}(\tau \hat{k}-\mu |s|)\nonumber\\
	=&-s^{T}\Theta s-\tilde{d}^{T}\left (\Lambda J^{-1} - \frac{1}{2}I_{3}\right)\tilde{d}+\frac{1}{2}\tilde{k}^{T}\tilde{k}\nonumber\\ 
	&+\hat{k}^{i}\tau_{i}\tilde{k}_{i}-|s|\mu \tilde{k}. \label{eq_sp_1}
    \end{align}
Using the fact that
\begin{align}
	\tilde{k}^{T}\mu |s| \leqslant \frac{\lambda_{\text{max}}(\mu)}{2} \left (\tilde{k}^{T}\tilde{k}+s^{T}s \right),
	\nonumber
\end{align}
(\ref{eq_sp_1}) can be expressed as
\begin{align}
	\nonumber
	\dot{V}\leqslant& -s^{T}\left(\Theta-\frac{1}{2}\lambda_{\text{max}}(\mu)I_{3}\right)s-\tilde{d}^{T}\left (\Lambda J^{-1}-\frac{1}{2}I_{3}\right)\tilde{d}\\ 
	\nonumber
	&+\frac{1}{2}\tilde{k}^{T}\tilde{k}+\hat{k}^{i}\tau_{i}\tilde{k}_{i}+\frac{\lambda_{\text{max}} (\mu)}{2}\tilde{k}^{T}\tilde{k}
	\nonumber
    \end{align}
    \begin{align}
	=& -s^{T}\left(\Theta-\frac{1}{2}\lambda_{\text{max}}(\mu)I_{3}\right)s-\tilde{d}^{T}\left (\Lambda J^{-1}-\frac{1}{2}I_{3}\right)\tilde{d}\nonumber\\
	&+\left (\frac{\lambda_{\text{max}}(\mu)+1}{2} \right ) \tilde{k}^{T}\tilde{k}+\tau^{i}\left(k_{i}\hat{k}_{i}-\hat{k_{i}}^{2}\right).
	\label{speq1}
\end{align}
Consider the following inequality
\begin{align}
	k_{i}\hat{k}_{i}-\hat{k}_{i}^2 &=-\left( \frac{k_{i}}{\sqrt{2}}-\frac{\hat{k}_{i}}{\sqrt{2}} \right)^{2}-\frac{\hat{k}_{i}^2}{2}+\frac{k_{i}^2}{2} \leqslant -\frac{\tilde{k}_{i}^{2}}{2}+\frac{k_{i}^2}{2}.\label{speq2}
\end{align}
Hence, by substituting the above \eqref{speq2} in equation \eqref{speq1}
\begin{align}
	\dot{V}\leqslant &-s^{T}\left(\Theta-\frac{\lambda_{\text{max}} (\mu)}{2}I_{3}\right)s-\tilde{d}^{T}\left (\Lambda J^{-1}-\frac{1}{2}I_{3}\right)\tilde{d}\nonumber\\
	\nonumber
	&-\frac{1}{2}\tilde{k}^{T}\left(\tau-\lambda_{\text{max}}(\mu)I_{3}-I_{3} \right)\tilde{k}+\frac{1}{2}k^{T}\tau k\\
	\implies \dot{V} \leqslant& -\Gamma V+ \bar{\Delta}, \label{lya_guub2}
\end{align}
where $\Gamma=\text{min}\left(\lambda_{\text{min}}(\Bar{\Theta}), \lambda_{\text{min}}(\Bar{\Lambda}),\lambda_{\text{min}}(\frac{\tau_{0}}{2}) \right)$, $\Bar{\Theta}=\Theta-\frac{\lambda_{\text{max}}(\mu)}{2}I_{3}$, $\Bar{\Lambda}=\Lambda J^{-1}-\frac{1}{2}I_{3}$ and $\bar{\Delta}=\lambda_{\text{max}}(\tau)||k||^{2}$. The equation (\ref{lya_guub2}) can be expressed as
\begin{align}
	\dot{V}\leqslant-\varrho V - (\Gamma-\varrho)V+\bar{\Delta}.
	\nonumber
\end{align}
For $\dot{V}(t) \leqslant 0$ for $V(0)> \mathcal{R}_{\text{sc}}$, there exists some $\mathcal{R}_{\text{sc}}=\frac{\Bar{\Delta}}{\Gamma - \varrho}$ such that $V(t)$ reaches the closed ball of radius $\mathcal{R}_{\text{sc}}$ in finite time, say $t_{r_{\text{sc}}}$ and remain there $\forall t >t_{r_{\text{sc}}}$. $t_{r_{\text{sc}}}$ can be evaluated as 
\begin{align}
	t_{r_{\text{sc}}}& \leqslant \frac{1}{\Gamma}\text{ln}\left[ \frac{V(0)-\frac{\bar{\Delta}}{\Gamma}}{\bar{\Delta}\left( \frac{1}{\Gamma-\varrho}-\frac{1}{\Gamma}\right)}\right].
	\nonumber
\end{align}
As the sliding variable component in the Lyapunov function, $\frac{1}{2}s^{2} \leqslant V $, the state trajectories are also contained within a closed ball of radius $\sqrt{\frac{2 \Bar{\Delta}}{\Gamma - \varrho}}$.
\begin{figure}[h!]
    \centering
    \subfloat[Evolution of quaternions\label{spacecraft_quaternions}]{%
        \includegraphics[width=0.23\textwidth,keepaspectratio]{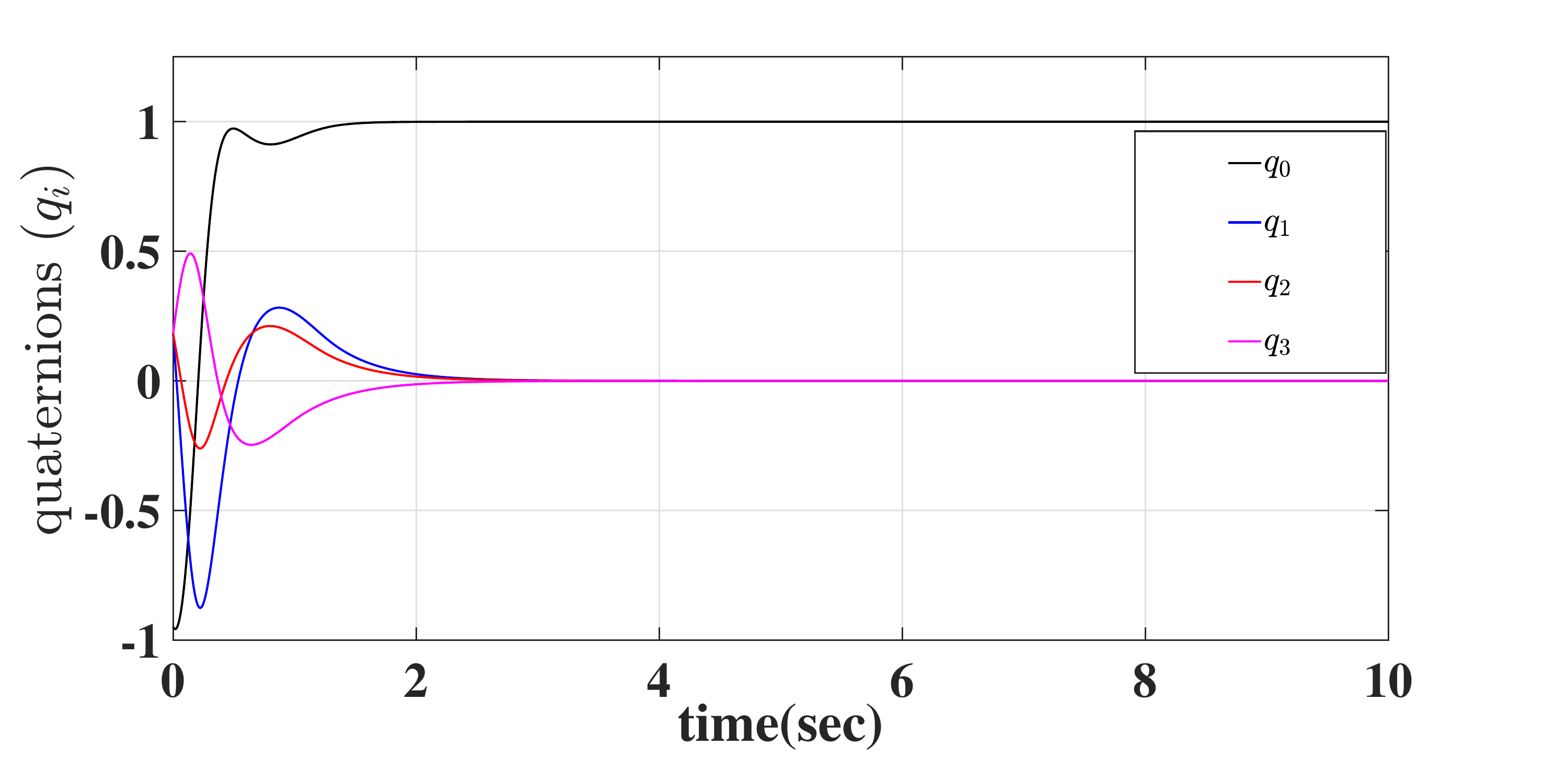}}
    \hfill 
    \subfloat[Evolution of angular velocities \label{spacecraft_omega}]{%
        \includegraphics[width=0.23\textwidth,keepaspectratio]{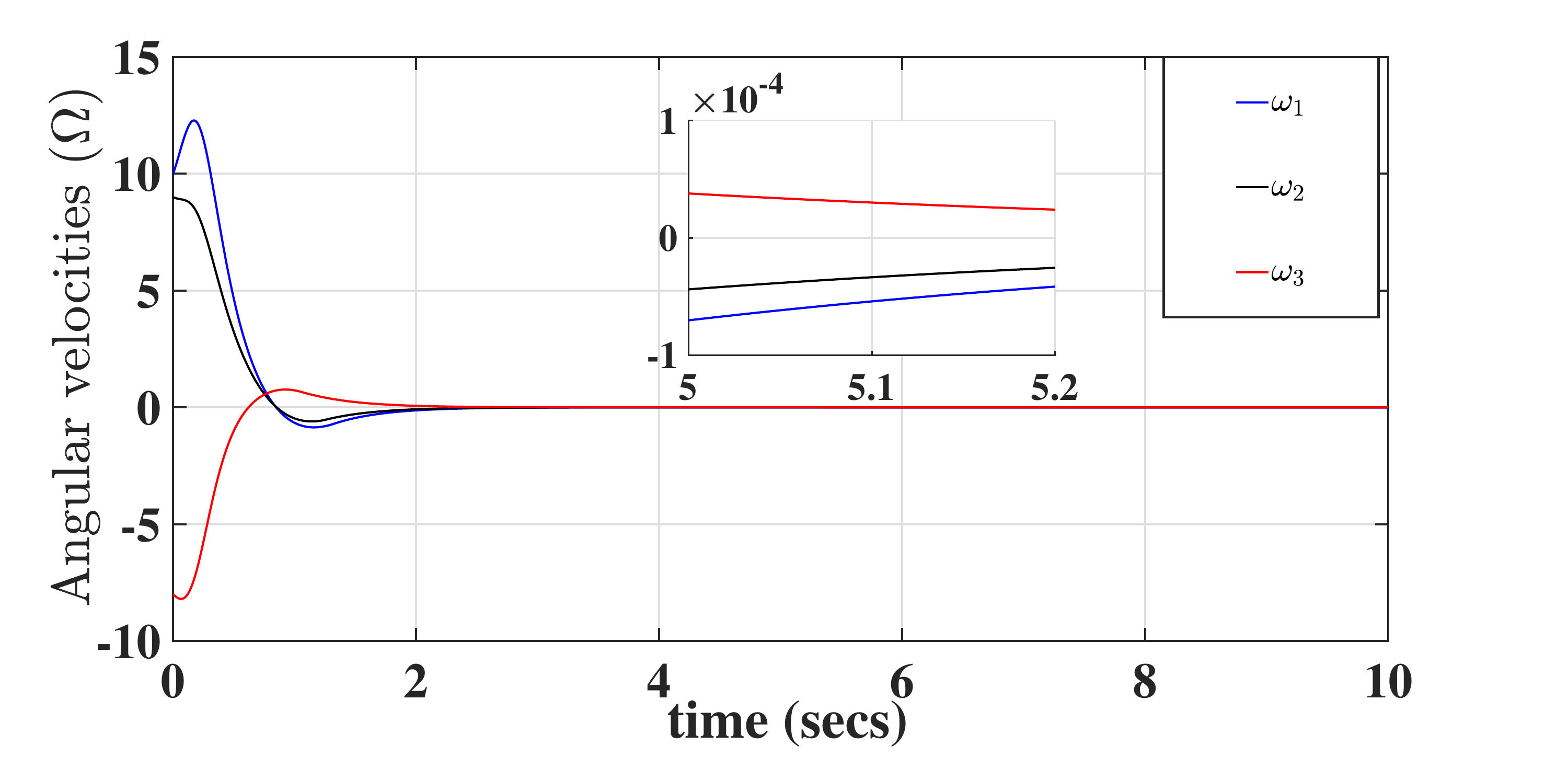}}
    \\
    \subfloat[Sliding variable of FOITSM for rigid spacecraft dynamics\label{spacecraft_ss}]{%
        \includegraphics[width=0.23\textwidth,keepaspectratio]{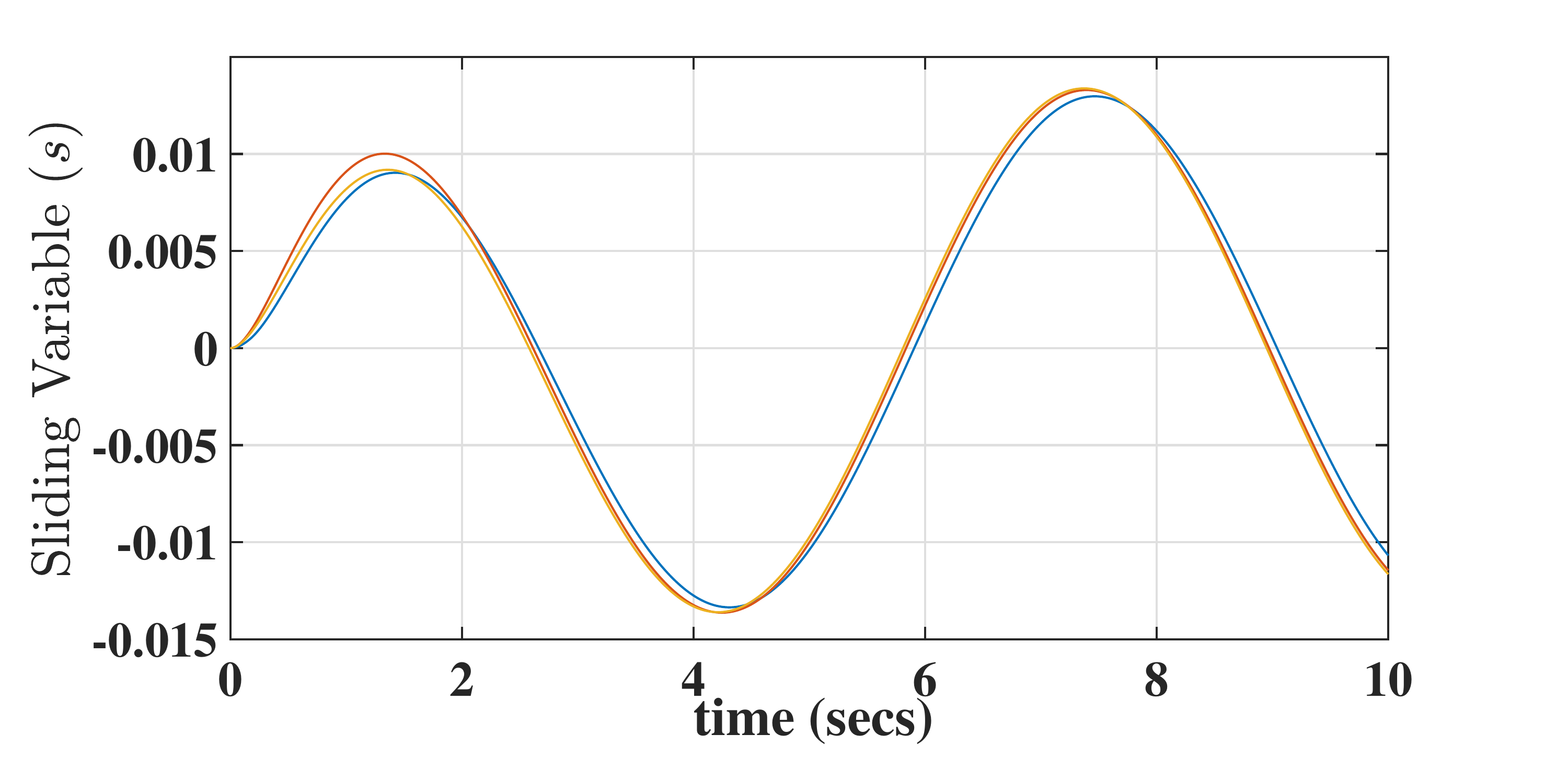}}
    \hfill 
    \subfloat[Disturbance estimation error of adaptive ADO with FOITSMC\label{spacecraft_d_tilde}]{%
        \includegraphics[width=0.23\textwidth,keepaspectratio]{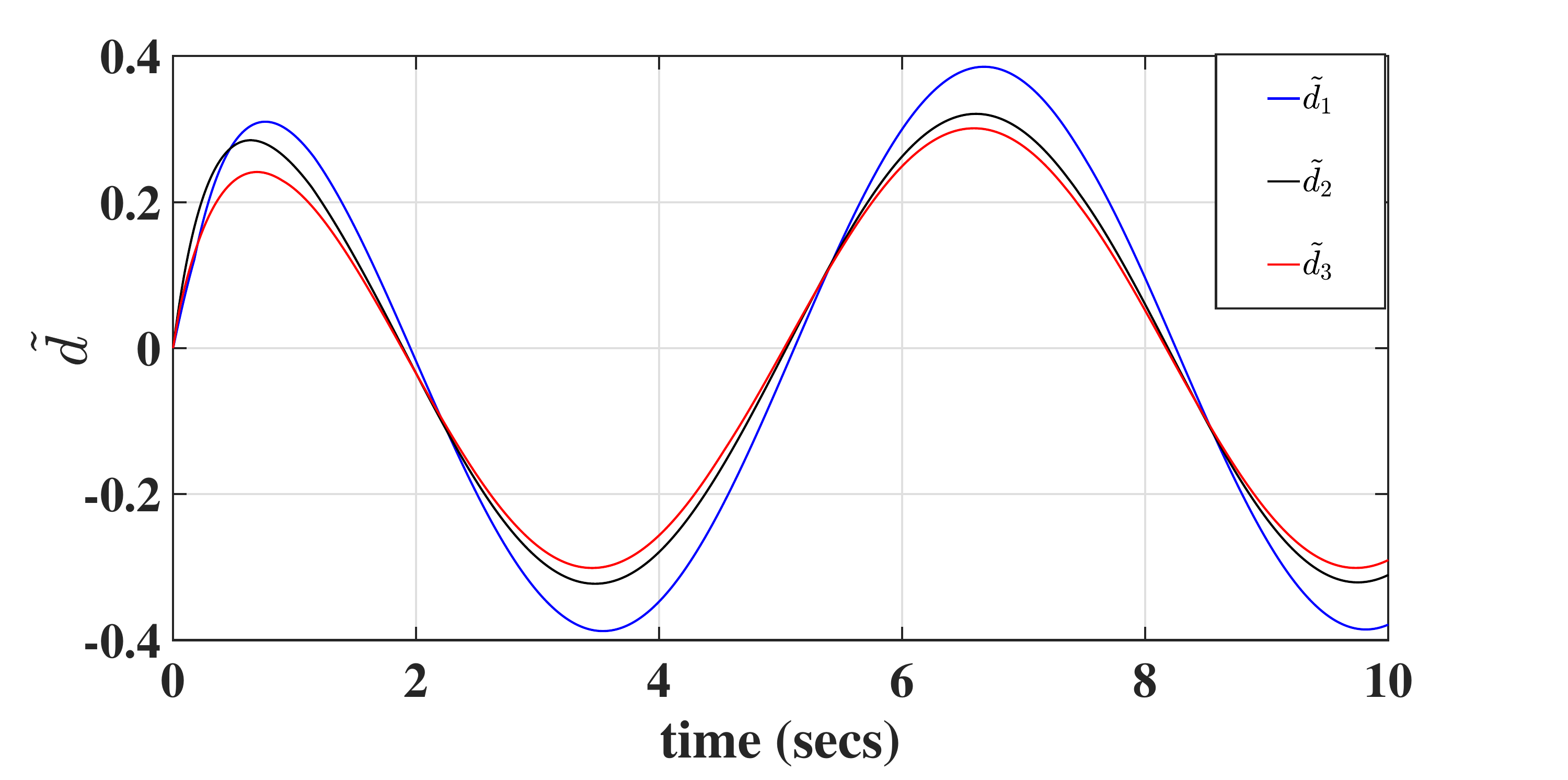}}
    \\
    \subfloat[Estimation of $k$\label{spacecraft_k_hat}]{%
        \includegraphics[width=0.23\textwidth,keepaspectratio]{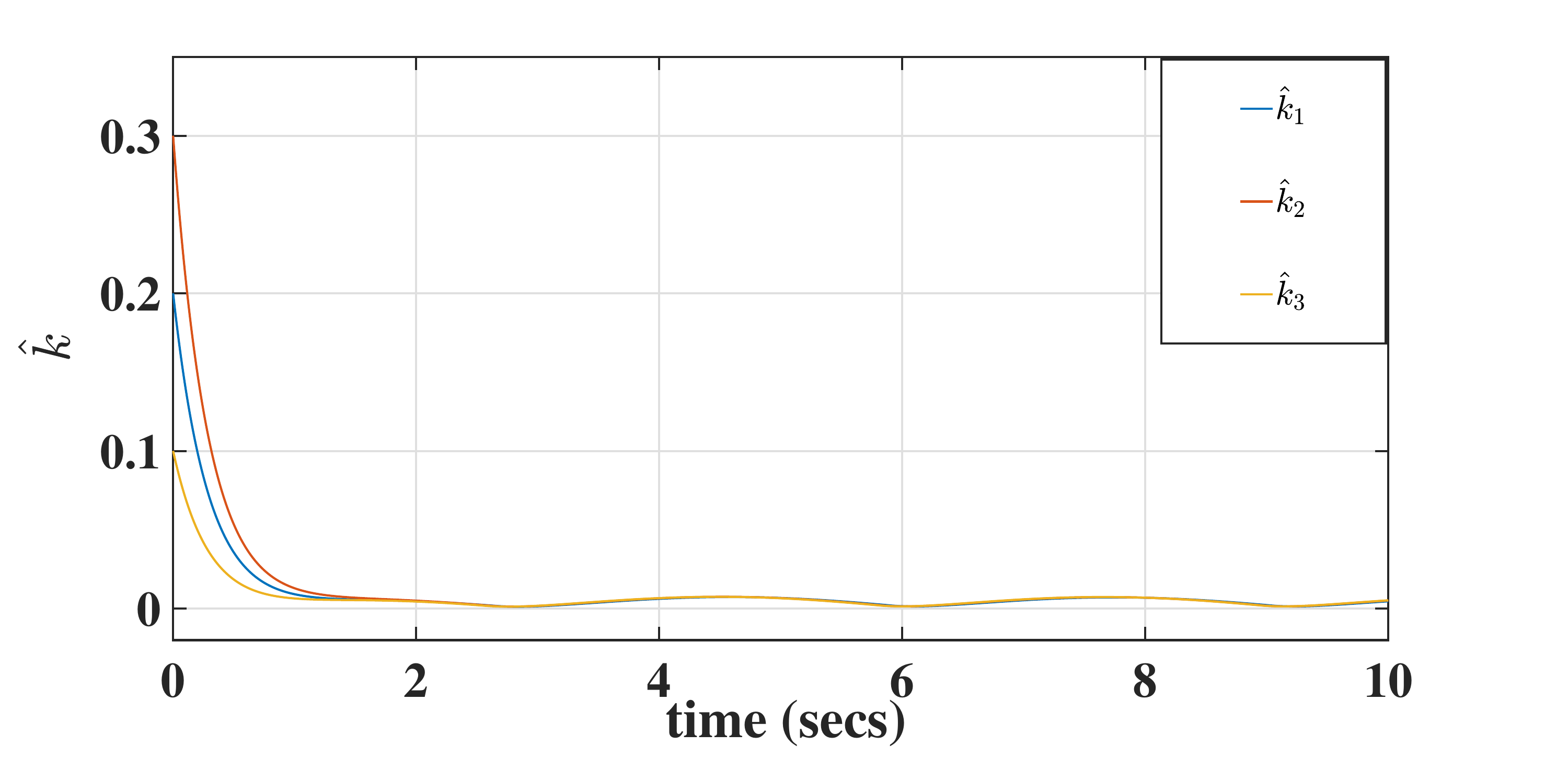}}
    \hfill 
    \subfloat[Input for FOITSMC with adaptive ADO\label{spacecraft_u}]{%
        \includegraphics[width=0.23\textwidth,keepaspectratio]{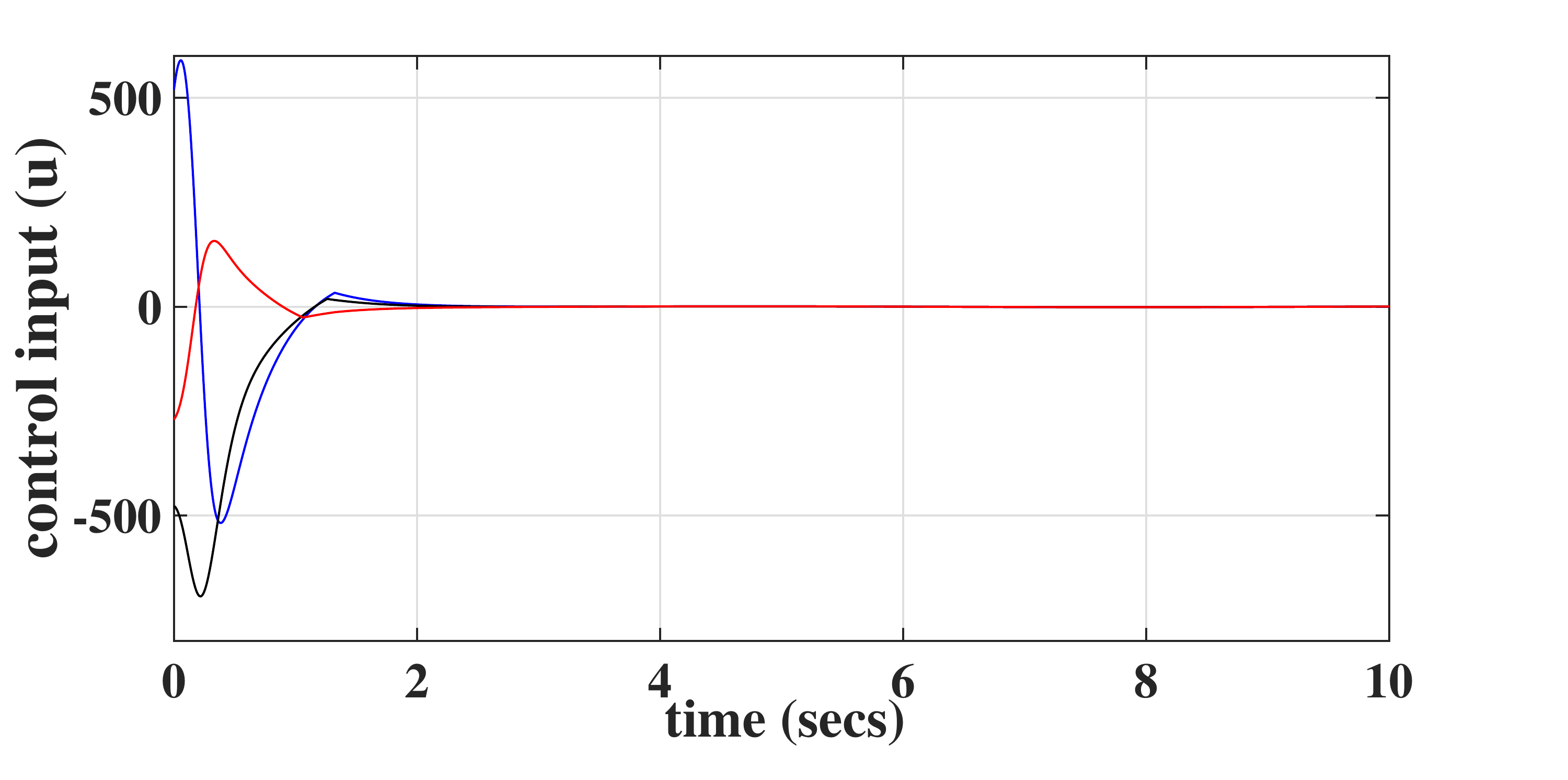}}
        
    \caption{\small This figure illustrates the performance of the proposed adaptive controller under matched disturbances. Sub-figures (\ref{spacecraft_quaternions})–(\ref{spacecraft_d_tilde}) confirm the global boundedness of the attitude quaternions, angular velocities ($\Omega$), the sliding variable, and the disturbance estimation error. The corresponding continuous control input and the evolution of the adaptive gain ($\hat{k}$) are presented in (\ref{spacecraft_u}) and (\ref{spacecraft_k_hat}), respectively.}
    \label{fig1_spacecraft}
\end{figure}
The parameters used in the simulation are as follows: $\Theta=2 I_{3}$, $\Lambda=50I_{3}$, $\mu=2I_{3}$, $\tau=5I_{3}$ and the inertia matrix,
\begin{align}
	J=\begin{pmatrix}
		20& 0 & 0.9\\
		0 & 17&  0 \\
		0.9& 0& 15
	\end{pmatrix}
	\nonumber
\end{align}

\par
The simulation results shown in fig. \ref{spacecraft_quaternions}  and \ref{spacecraft_omega}, utilising the ODE 4 (Runge-Kutta) solver of MATLAB R2021a with an assigned fixed step size of 0.001, depict that the state trajectories remain close to the origin, bounded within a specified limit. A continuous control input has been employed using an adaptive ADO-based FOITSMC algorithm, as shown in Fig. \ref{spacecraft_u}, to ensure the global boundedness of error in the estimation of conditionally known disturbance and unknown gain $k$.

\bibliographystyle{IEEEtran}
\bibliography{IEEEabrv,paperLetters.bib}

\end{document}